\documentclass[preprint,12pt]{elsarticle}
\usepackage{amsmath, mathrsfs, latexsym, amssymb, color, graphicx, comment, hyperref, scrhack}
\usepackage{lmodern}
\usepackage{physics}

\usepackage{amssymb}

\newcommand{\dfr}[2]{\frac {\displaystyle #1}{\displaystyle #2}}
\journal{Physica A}

\begin{document}

\begin{frontmatter}

\title{Description of Glass Transition kinetics in 3D $\mathrm{XY}$ model in terms of Gauge Field Theory}


\author[label1,label2]{M.\,G.\,Vasin}
\address[label1]{Institute for High Pressure Physics of Russian Academy of Sciences, 142190 Moscow, Russia}
\address[label2]{Udmurt Federal Research Centre of UrB Russian Academy of Sciences, 426067 Izhevsk, Russia}

\begin{abstract}
We consider a gauge theory of the glass transition in the frustrated XY model being simplest model containing topologically nontrivial excitations. We describe the transition kinetics and find that the three-dimensional system exhibits the Vogel--Fulcher--Tamman criticality heralding its freezing into a spin glass. We analytically show that the system demonstrates all glass transition properties, like the logarithmic relaxation, and corresponding behavior of linear and non-linear susceptibility. The mode-coupling theory equation in the Zwanziger--Mori representation also is derived in framework of our approach. Our findings provide insights into the topological origin of glass formation, that allows to make progress in understanding glass-transition processes in more intricate systems.
\end{abstract}


\end{frontmatter}


\section{Introduction}
\label{S1}

The formulation of a universal theory of glass transition remains for a long time one of the most intriguing but still unresolved problem of condensed matter physics~\cite{N1, N2, Tanaka}. Large number of systems, which manifest this phenomena, allows us to conclude regardless of their nature  that this phenomena does not depend on microscopic details, but it is determined  by the symmetry properties of systems independently of the scale, like in the case of phase transitions. Besides we know that during the glass transition the critical slowing down  takes place of all processes at the microscopic level the same as at second order phase transition. However, it is well known that the glass transition is not accompanied by the characteristic for critical phenomena divergence of the static correlation length of the order parameter and the susceptibility that does not allow us to assign the glass transition to the phase transition. And more an order parameter just is absent because of absence any ordering. Furthermore, the key feature of nonequilibrium strongly disordered systems is the presence of the broad spectrum of relaxation times. As the result it is challenge to write any dynamical (kinetic) equations for glassy systems since it is quite complicated to divide the scales on ``fast'' and ``slow'' ones.

The simplest explanation why liquids can freeze into a glass rather than to crystallize upon the quench follows from that the rapidly growing and then colliding nuclei of the solid phase may have different orientations. The multiple topological defects then appear at the interfaces between crystallites, and it takes exponentially long times to anneal them to establish the long-range crystalline order. This mechanism resembles the celebrated Kibble-Z\"{u}rek scenario of the universe formation~\cite{Kibble,Zurek}. However, that liquid relaxation time can diverge according to different scenarios indicates that the systems makes its freezing choice well in advance. One would expect a somewhat different solidification scenario if the nucleation starts already far in the liquid phase, at the temperatures below the so called Frenkel line~\cite{Brazhkin1,Brazhkin2} (named in honor of Jacob A. Frenkel), when the orientation hardness appears in liquid. Then the system's structure becomes nonhomogeneous because of the topological perturbations presence.
There are handsome theoretical approaches that suppose the topological perturbations play central role at the glass transition.
Then the glass transition has the same topological nature as the Berezinskii-Kosterlitz-Thouless (BKT) phase transition in two-dimension systems, and should be described within non-perturbative methods of quantum field theory.

The BKT transition is the binding--unbinding transition between the low-temperature phase, $T<T_{BKT}$ ,where the topological excitations (vortices) of the opposite sign are bound into the `neutral' dipoles, and the high-temperature unbound phase, at $T>T_{BKT}$, where topological excitations unbind loose and form a `free' neutral plasma\,\cite{Ber,KT1972,KostThoul}.
Among many remarkable properties of the BKT transition, the singularity of its critical behavior stands out. On approach to $T_{BKT}$ from the above, the correlation length that sets the spatial scale for separation between the free excitation, diverges much faster than any power law governing the correlation length for a standard continuous phase transition\,\cite{Kosterlitz1977,Jose1977}.
This BKT criticality strikingly resembles one, which is observed near the glass transition with the relaxation time diverging according to the Vogel--Fulcher--Tammann (VFT) law.
The possible topological nature of VFT criticalities was indicated by Anderson\,\cite{Anderson1978}, who attributed the VFT criticality to logarithmic interaction between the topological excitations. In \cite{VV} we have proved this proposition on the example of simple model.

Addressing to the description of the vitrification kinetics, we can note that, despite the presence of short-range ordering in glass systems, the growth of ordered regions is impeded by the blocking of their motion, caused either by a frozen disorder presence or by the intrinsic topological nature of the system. In both cases one can talk of the presence of topologically stable perturbations in the structure, frustrating the system. The concept of the topologically stable perturbations has long been used for the theoretical description of spin, vortex, quantum glasses, and amorphous substances. The term of "frustration" was introduced by Toulouse and Villain from a microscopic point of view, as a topological property of lattices with opposite signs bonds. The Ising and X-Y models in two and three dimensions in a frozen distribution of frustrations was studied by Fradkin \cite{F1,F2}. The macroscopic interpretation of frustration in the XY model and the Heisenberg model was obtained by Volovik and Dzyaloshinskii \cite{R37, R38}. From the macroscopic point of view the microscopic frustration lines of Toulouse and Villain \cite{R22, R39} are the topologically stable perturbations (topological defects), on which the rotational symmetry of the initial local structure of matter is violated \cite{N1,R23,R34,R40}.

The topological phase transition associated with the appearance of topologically stable linear perturbations (topological defects) of infinite length is well known as the crystal melting model \cite{R42,Obukhov}. Inherently, the transition to the glass state from the high-temperature liquid phase is also a topological phase transition. But between the topological melting and the glass transition there is a significant difference. In the first case, the topologically stable perturbations arise as a result of thermal excitation, in glasses these perturbations exist at low temperatures, in a state with a frozen configuration \cite{R37}.

In the present work we focus on the XY model representing a wealth of physical systems ranging from Josephson junction arrays and vortex systems in type II superconductors to spin glasses that exhibit glassy behavior. In two dimensions, the XY model is a generic system for the BKT transition. At the same time, the 3D XY model has been for decades an exemplary testing ground for studying a glass transition in spin systems, see, for example early papers\,\cite{R38,DV5}.

The glass formation in XY systems has been a subject of tireless attacks based on the quantum field theory methods\,\cite{R38,R23,DV5,Hertz} and computational approaches, see, for example,\,\cite{R44,Olsson1,Olsson2,R45,R46}. Nevertheless, in spite of the expended substantial efforts, the details of the vitrification process and its relation to topological excitations in XY systems still remain a mystery.
In \cite{VV} we have developed a gauge theory of the topological transition in the XY model. Our approach follows the ideas proposed by Rivier and Dzyaloshinskii \cite{R38,R23,DV5,Hertz}, and describes the system in the terms of gauge fields induced by mobile vortices.
We had shown that the transition in frustrated 3D XY-model had the same topological nature as BKT transition, and has distinctive cues of glass transition. In the present paper we are expanding the above approach, describing the 3D system dynamics close to this transition.

\section{Model}
\label{S2}

First of all let us consider pure, i.e. without any frustrations, 3D XY model on a lattice. It is a 3D grid with the two-component classical vector of the unit length ${\bf S_r}=(\cos \varphi_{{\bf r}},\,\sin\varphi_{{\bf r}})$ assigned to every nod ${\bf r}$. Each vector can rotate in the XY plane.
The system's Hamiltonian is:
\begin{equation*}
H=-\dfr{1}2\sum\limits_{\langle {\bf r}\neq{\bf r'}\rangle }^N\mathcal{E}_{\bf r-r'}{\bf S_r}{\bf S_{r'}}=-\dfr{\mathcal{E}}2\sum\limits_{\langle {\bf r}\neq{\bf r'}\rangle }^N\cos\left(\varphi_{\bf r}-\varphi_{\bf r'}\right).
\label{H1}
\end{equation*}
where $\mathcal{E}$ is the coupling energy of nearest vectors,
$N$ is the total number of the nodes, and brackets  $\langle {\bf r}\neq{\bf r'}\rangle$ stand for the summation over the nearest neighbors around ${\bf r}$. Hereinafter we will be interested in Weiss theory in the long wave limit, and it is well known\cite{Polyakov} that in the continuous limit this theory is equivalent to the theory with the following Hamiltonian:
\begin{gather*}\label{L1}
H_0=\dfr{\epsilon}2|\nabla\Psi|^2-\dfr{m^2}2|\Psi|^2+\dfr b4|\Psi|^4,
\end{gather*}
where the vector field $\Psi_{\bf r}=\langle {\bf S}_{\bf r} \rangle_{\delta V}$ is convenient to be represented in the complex form $\Psi_{\bf r}=\psi_{\bf r} e^{i\varphi_{\bf r} } $. This field coarse grained over some finite volume $\delta V$ around the space point ${\bf r}$, plays the role of the Higgs field, $b>0$, $m^2=\alpha (T_c-T)/T$ with $\alpha >0$, and $T_c$ is the temperature at which the symmetry breaking of the system's order parameter happens, $\langle \Psi \rangle=0 \to \langle \Psi \rangle\neq 0$. Note that in contradistinction to 2D system, where the thermal fluctuations destroy long range order, in 3D system does undergo second order phase transition at $T_c$. However, as we will see below, the frustration can markedly transform this transition.

Let us consider the above system with frustration. Usually, the frustration is introduced in the model through randomness included through coupling constants. This randomness is expressed through an random gauge field $A$:
\begin{equation*}
H=-\dfr{1}2\sum\limits_{\langle {\bf r}\neq{\bf r'}\rangle }^N\mathcal{E}_{\bf r-r'}{\bf S_r}{\bf S_{r'}}=-\dfr{\mathcal{E}}2\sum\limits_{\langle {\bf r}\neq{\bf r'}\rangle }^N\cos\left(\varphi_{\bf r}-\varphi_{\bf r'}+A_{\bf r-r'}\right).
\label{H2}
\end{equation*}
Thus the frustration existence in the system defines the gauge field presence in the theoretical description of this system.
In order to take it into account in continuous model, one introduces the covariant derivative instead of ordinary one, $\nabla \to D=\nabla -ig{\bf A}$, where $g$ is the coupling constant\,\cite{R38,Hertz,R2}. Then, believing the gauge field is also continuous, one can write the Hamiltonian in the following form:
\begin{gather}\label{L2}
H_0=\dfr{\epsilon}2|D\Psi|^2+\dfr{\kappa}2(\nabla\times{\bf A})^2-\dfr{m^2}2|\Psi|^2+\dfr b4|\Psi|^4.
\end{gather}
Since we are interested in not yet freeze system, at variance to the approach adopted by Hertz\,\cite{Hertz} we suppose that the gauge field is not constant, and the correlator $\langle {\bf AA}\rangle$ can be determined self-consistently.
In the low-temperature state, $T<T_c$, the equilibrium $|\Psi|^2$ is non-zero, $|\Psi|^2=\psi^2=\alpha (T_c-T)/Tb$, at that Eq.\,(\ref{L2}) has non-trivial solutions corresponding to the topologically protected vortex excitations.
As we noted above, according to works of Toulouse, Villain and Dzyaloshinskii \textit{et al.}\,\cite{R38,R22,R39,DV5}, the frustrations correspond to the  topologically stable vortices which are the gauge field sources.
We think that frustration degree defines the maximal vortices number (or their maximal concentration, $\lambda_0$, proportional to the number of the quenched disordering points). Then the corresponding Hamiltonian for the system with $N$ vortices has the following form:
\begin{gather*}
H_N=\mathcal{H}_0+i\sum\limits_{n=1}^N{\bf J}_n{\bf A}=\dfr{\kappa}2{\bf A}\nabla^2{\bf A}-\dfr{M_0^2}2{\bf A}^2 +i\sum\limits_{n=1}^N{\bf J}_n{\bf A},
\label{H5}
\end{gather*}
where $\bf J$ denotes the `elemental current' or `elemental charge' associated with the single vortex, $\bf A$ is the gauge field mediating the interaction of these vortices, and $M_0^2=\epsilon g^2\alpha (T_c -T)/Tb$ is the square of gauge field mass appearing due to of Anderson--Higgs mechanism, i.e. the Higgs mass.
In the momentum representation the gauge field Green function is
\begin{gather*}\label{Field}
    \langle {\bf A}({\bf p}){\bf A}(-{\bf p})\rangle_A=\Delta ({\bf p})=\dfr{-\beta^{-1}}{{\kappa}{\bf p}^2+M_0^2}\,,
\end{gather*}
where $\langle \ldots \rangle_A$ denotes the statistical averaging over all possible configurations of $\bf A$ field, $a$ is the vortex core size, and $\beta=1/k_BT$ ($k_B$ is the Boltzmann constant).
The characteristic correlation length scale is proportional to the inverse Higgs mass, $\xi \sim M_0^{-1}$.
One immediately sees that vortices behave like currents in magnetostatics.
When the mass of the gauge field turns zero, the system undergoes phase transition. In the absence of vortices, this transition occurs at $T=T_c$, but formation of vortices shifts the transition.

\section{Averaging over grand canonical ensemble of vortices}

In order to take account of all possible vortex configurations, we utilize the Grand Canonical Ensemble description of the vortex gas. Mobile vortices screen each other giving rise to the renormalization of the gauge field mass. To calculate the renormalized mass, one averages over the grand canonical ensemble of the ``particles'' endowed with the two possible dimensionless charges.
The partition function of the vortex ensemble is:
\begin{multline*}
Z=\langle e^{ -\beta H}\rangle=\langle\langle e^{ -\beta H}\rangle_A\rangle_J=\left<\int\mathcal{D}{\bf A}
e^{ -\beta a^{-d}\int \mathrm{d}^d {\bf r}H_N}\right>_J \\
=\sum\limits_{N=1}^{\infty}\dfr {\lambda^N}{N!}\sum\limits_{\{J_n\}}\int\mathcal{D}{\bf A}
\int\prod\limits_{n=1}^N \mathrm{d}^d{\bf r}_ne^{ -\beta a^{-d}\int \mathrm{d}^d {\bf r}\left[H_0-i\delta^{(2)}({\bf r}-{\bf r}_n){\bf J}_n({\bf r}){\bf A}({\bf r})\right]},
\end{multline*}
where $\langle \ldots \rangle_J$ denotes the statistical averaging over grand canonical distribution of vortices, $\{{\bf J}_n\}$ is the set of all configurations of ${\bf J}_n$, $\lambda=\lambda_0e^{-\beta E_{c}}$ is the non-dimensional factor proportional to equilibrium vortices number,
$\lambda_0$ is the frustration density, proportional to the number of the quenched disordering points, and $E_c$ is the vortex core energy. As we will see, parameter $\lambda_0$ controls the kinetics of the system upon cooling.

Note that according to the topological laws in 3D system with SO(2) symmetry the vortex is the linear. Therefore, from the least action principle, one derives the part of the action containing the gauge field source:
\begin{gather*}
i\beta a^{-d}\int \mathrm{d}^d{\bf r} \delta^{(2)} ({\bf r}-{\bf r}_n){\bf J}_n{\bf A}({\bf r})\approx i\beta J_n|{\bf A}({\bf r}_n)|,
\end{gather*}
where ${\bf J}_n$ is the topological source.
Then averaging over the dimensionless quantity ${\bf J}_n=\pm J$ one arrives at the
\begin{multline*}
Z=\int\mathcal{D}{\bf A} e^{-\beta a^{-d}\int\mathrm{d}^d{\bf r}H_0}\sum\limits_{N=1}^{\infty}\dfr {1}{N!}\left(2\lambda\int \mathrm{d}^d{\bf r}\cos\left[ \beta J|{\bf A}({\bf r})|\right]\right)^N \\
=\int\mathcal{D}{\bf A} \exp\left(-\beta a^{-d}\int \mathrm{d}^d{\bf r} \left( H_0 -2\lambda\beta^{-1}\cos\left[\beta J|{\bf A}({\bf r})|\right]\right)\right).
\end{multline*}
Note, that the averaging is carried out over all quantities of the point vortices and their possible positions.  When averaging over all the configuration of the 3D system with the proper weight, the protocol automatically ``chooses'' only linear configurations of the vortices. Thus, this averaging takes into account all configurations of the linear vortices including all possible loop configurations.
As a result the system's effective Hamiltonian becomes:
\begin{gather}
H=\dfr{\kappa}2{\bf A}\nabla^2{\bf A}-\dfr{M_0^2}2{\bf A}^2-2\lambda\beta^{-1} \cos\left(\beta J|{\bf A}|\right)\,,
\label{Z1}
\end{gather}
which is nothing but the Hamiltonian density of the sine-Gordon theory\,\cite{Min}.

The quadratic term of the power series expansion of cosine in Eq.\,(\ref{Z1}) renormalizes the Higgs mass, $M^2_0\to M^2\approx \epsilon{g}^2\alpha (T_c-T)/Tb-2\lambda J^2/k_BT$ reflecting the loss of the system's spin collinearity of vortices at $T<T_c$. Accordingly, the critical temperature for the gauge field, determined by the condition $ \delta^2 \mathcal{H}/ \delta A^2|_{A=0}=0$, shifts from $T_c$ to $T_g=T_c-2\lambda bJ^2/k_B\epsilon{g}^2\alpha $.
Eventually, one concludes that in the $T_c>T>T_g$ temperature interval the system falls into the state that has local ordering, but which does not possess the long-range order since the latter is destroyed by mobile vortices. This state is referred to as {\it disordered phase}. At $T=T_g$ the system  undergoes a phase transition, the features of which depend on the dimensionality of the system.
The peculiarity of this transition is that it is the topological phase transition in which an order parameter does not arise at $T_g$ but the correlation radius of the gauge field diverges, the field becomes massive, and, as a result, at $T<T_g$ the system freezes into the state, which is called as {\it confined phase}\,\cite{Polyakov,Zee}.

The above expressions hold in the general $d$-dimensional case. In 2D systems with the degenerate continuous symmetry, the topological phase transition is the Berezinskii--Kosterlitz--Thouless transition. In 3D systems the entropy of a linear vortex increases linearly and so does the associated elastic energy. Therefore, in the 3D system vortices can emerge and become relevant in the presence of the additional perturbation caused by the quenched disorder frustrating the system. If in the 3D case topological excitations were absent, then the mass of the vector field $M=M_0$ would have become zero at $T=T_c$. Then the behavior of the XY-system would not differ from that of the Ginsburg--Landau system, and $T_c$ would have become the temperature of the second order phase transition. The appearance of the statistically relevant number of topological excitations leads to the renormalization of the mass of the gauge field and to the emergence of disordered phase.

We should note, that if we consider a system with low vortices density, when the frustration does not eliminate the second order phase transition. The phase transition does occur, but at the temperature $T<T_c$.
At $T=T_c$,  the nonzero local magnetization arises, $|\Psi|^2=\psi^2$, but the frustration generates vortices that destroy the order on large scales. As a result, in some temperature interval below $T_c$ the phase transition still does not occur, since in this temperature interval the correlation radius, $r_c$, of the correlation function $\langle \Psi\Psi\rangle_{\bf r}$ remains finite.
In other words, the low density vortices renormalizes the phase transition temperature, and shifts it downwards, $T_c\to T_c^R$.
This renormalized phase transition temperature depends on the vortex concentration and can be estimated in the one-loop approximation \cite{VV}:
\begin{gather*}
T_c^R(\lambda )=T_c-\dfr{T_c\epsilon g^2a\kappa^{1/2}}{\alpha}\sqrt{2\lambda J^2/k_BT_c}.
\end{gather*}
In contrast to the glass transition, here the diverging quantity is the order parameter correlation length.
Thus, if at some vortex density $\lambda $ the inequality $T_c^R(\lambda)>T_g(\lambda)$ holds, then the system experiences the second order phase transition, at $T=T_c^R$. In the opposite case,  $T_c^R(\lambda)<T_g(\lambda)$, the system undergoes the transition to confined phase (glass transition) at $T=T_g$ \cite{VV} .

\section{Renormalization of the vector field mass}

Now we focus on the critical behaviors specific to a particular. Let us expand the cosine in the Hamiltonian density in the Taylor series over the powers of $A$.
However, the quantum field theory\,\cite{Zee} teaches us that in 3D case only terms with the power of $A$ that is less than $n=6$ ($n < 2d/(d-2)$) are relevant in the Taylor expansion\,\cite{Zee,Cvel}. Furthermore, it is known\,\cite{Polyakov} that the remaining effective nonlinearity is exponentially small, $\beta\lambda \sim e^{-\beta E_c}\ll 1$, and is described by the Debye approximation. In this case the Debye volume, $V_D$, contains sufficiently many particles in order to neglect the fluctuations of the sum of their fields.
Indeed, since the particle density $\lambda \propto e^{-\beta E_c}$, then from (\ref{Z1}) the particle number in the Debye volume, $V_D\sim(M_0^{2}+a\lambda)^{-3/2}\propto e^{3\beta E_c/2}$, exponentially diverges: $\lambda V_D\sim e^{\beta E_c/2}\gg 1$.
It implies that at $d>2$ the perturbation series of the sine-Gordon theory do not contain infrared divergences\,\cite{Cvel}. Hence the system's Hamiltonian assumes the form
\begin{gather*}
H=\dfr{\kappa}{2}{\bf A}\nabla^2{\bf A}-\dfr{M^2}{2}{\bf A}^2,
\end{gather*}
where
\begin{gather*}
M^2=\epsilon g^2\alpha (T_{g}-T)/Tb.
\label{ME5}
\end{gather*}

In \cite{VV} we shown that the confinement at $T=T_g$ manifests main features of the glass transition. In particular
a straightforward evidence for the glassiness follows from the behavior of the spin correlation function. Stable vortices destroy long-range order in the low-temperature phase that forms at $T<T_g$, i.e. $\langle \Psi \rangle =0$. Accordingly, spin correlation function $\langle\Psi\Psi\rangle|_{r\to\infty}=0$, and the spin correlation length is small, $\sim m^{-1}=\sqrt{g/2\lambda b a^{d-1}}$.
The further evidence of the glass transition at $T=T_g$ in a spin system can be drawn from the behaviors of the linear and nonlinear susceptibilities on approach to the supposed transition temperature\,\cite{binder}. It was shown that the linear susceptibility of the system $\chi=\left.\partial \langle\Psi\rangle/\partial {\bf h}\right|_{h\to 0,\,p\to 0}=\beta\left.,\langle\Psi^2\rangle\right|_{p\to 0}$ (where $h$ is an external source of the field $\Psi $), was finite in $T_g$. This satisfies to the glass transition, unlike the infinitely divergent value at the second order phase transition. The nonlinear susceptibility is $\chi_N=\left.\partial^3\langle \Psi\rangle/\partial {\bf h}^3\right|_{h\to 0,\,p\to 0}=\beta^3\langle \Psi^4\rangle
_{p=0}$. We shown that close to $T_g$ it could be estimated as $\chi_N\propto -\ln(T-T_g)$ for $T\to T^{+}_g$.
This value diverges at $T=T_g$, that also corresponds to the glass transitions in the spin systems~\cite{binder}.
Hence the transition at $T=T_g$ belongs in the same universality class as the glass transition in the elastic media\,\cite{R40}.

Thus, both, the fact that linear susceptibility remains finite and the non-linear susceptibility diverges at $T\to T_g$, and that the long-range order is destroyed below $T_g$, $\langle \Psi \rangle =0$,
suggests strongly that $T_g$ is the glass transition temperature, see\,\cite{R23,R34,R2,binder,K2,Nussinov,Vasin}.
The considered physical picture also agrees with the frustration-limited domain theory~\cite{R34,K2} and with the gauge theory of glass\,\cite{R23,R2}.
Besides, earlier, using the non-equilibrium critical dynamics methods, we have shown that the frustrated 3D system, which undergoing the second order phase transition or weak first order phase transition, does not reach the low-temperature ordered state, but freezes  in the non-ergodic glass state~\cite{SupVasin}. Therefore, we conclude that $T_g$ is indeed the glass transition temperature.

\section{Dynamics}

We suppose  that near the glass transition the observable kinetic properties of the system are conditioned primarily by the vector field kinetics.
Therefore, in order to estimate the relaxation time of the system close to the glass transition we neglect by the Higgs mode.
Let us consider a system in which we distinguish the isolate vortex, ${\bf J}({\bf p},\,\omega )$, with fixed impulse $\{{\bf p}, \omega\}$.
We investigate the kinetics of the considered model near to the critical point $M=0$. To do this, we will consider the non-equilibrium dynamics of the system.  The usage of the functional technique for the description of non-equilibrium dynamics~\cite{Kamenev,HH} leads to the representation of the partition function of the system in the following form:
\begin{multline*}
\displaystyle \mathcal{Z}=\int D\vec{\bf A}\sum\limits_{{\bf J}=\pm J}\exp\left[ \mathcal{S}_A+
i\beta\bar{\bf A}({\bf p},\omega){\bf J}(-{\bf p},-\omega)+i\beta{\bf A}({\bf p},\omega)\bar{\bf J}(-{\bf p},-\omega)\right],
\end{multline*}
where
\begin{multline}\label{ActionA}
\displaystyle \mathcal{S}_A=\tau a^{d}\beta\int\mathrm{d}\omega'\mathrm{d}{\bf p}'\left[{\bf A}\left(-\kappa\nabla^2+\Gamma_A \tau\partial_t+M^2\right)\bar{\bf A}+\right.\\\left.
\bar{\bf A}\left(-\kappa\nabla^2-\Gamma_A\tau \partial_t+M^2\right){\bf A}-2\beta^{-1}{\bar{\bf A}}^2\right],
\end{multline}
$\vec {\bf A}=\left\{ \bar {\bf A},\,{\bf A}\right\}$ and $\vec {\bf J}=\left\{ \bar {\bf J},\,{\bf J}\right\}$ are vectors, the components of which are named ``quantum'' and ``classical'' respectively, $\tau $ is the characteristic relaxation time, $\Gamma_A \tau $ is the kinetic coefficient~\cite{Kamenev,HH}, which corresponds to the gauge field.
The Green functions of the massive vector field components in impulse form are:
\begin{multline}\label{GFA}
    \Delta^R_0({\bf p},\,\omega )=\langle \bar{\bf A}{\bf A}\rangle_{{\bf p},\,\omega }=\dfr{-\beta^{-1}}{\kappa{\bf p}^2+M^2+ i\Gamma_A\tau\omega },
    \\
    \Delta^A_0({\bf p},\,\omega )=\langle \bar{\bf A}{\bf A}\rangle_{{\bf p},\,\omega }=\dfr{-\beta^{-1}}{\kappa{\bf p}^2+M^2- i\Gamma_A\tau\omega },
    \\
    \Delta^K_0({\bf p},\,\omega )=\langle {\bf A A}\rangle_{{\bf p},\,\omega }=\dfr{2\beta^{-2}}{(\kappa{\bf p}^2+M^2)^2+\Gamma_A^2\tau^2\omega^2 }.
\end{multline}
These Green functions satisfy to the fluctuation-dissipation theorem:
\begin{gather*}
\,\Im\, \Delta^R({\bf p},\,\omega )=\dfr{\Gamma_{A}\tau\omega}{2k_BT}\Delta^K({\bf p},\,\omega ).
\end{gather*}

Using these functions we can write the vortex correlation function in the following form:
\begin{multline*}
\displaystyle \langle {\bf J}\bar{\bf J}\rangle_{{\bf p},\,\omega}=\left\langle{\bf J}_{{\bf p},\,\omega}\bar{\bf J}_{-{\bf p},\,-\omega}\exp\left[ \mathcal{S}_A+i\beta\bar{\bf A}({\bf p},\omega){\bf J}(-{\bf p},-\omega)+\right.\right.\\\left.\left.
i\beta{\bf A}({\bf p},\omega)\bar{\bf J}(-{\bf p},-\omega)\right]\right\rangle =
\exp\left[J^2\beta^2\langle\bar{\bf A}{\bf A}\rangle_{{\bf p},\,\omega}\right].
\end{multline*}
Thus
\begin{gather*}
\displaystyle \langle {\bf J}\bar{\bf J}\rangle_{{\bf p},\,\omega}\propto\exp\left[ \dfr{-\beta J^2}{\kappa{\bf p}^2+M^2+i\Gamma_A\tau\omega } \right].
\end{gather*}

The vortex correlation function decays exponentially with time $t$,  $G(t)=\langle{\bf J}\bar{\bf J}\rangle_t\propto \exp(-t/\tau_r)$, where $\tau_r$ is the relaxation time.
Its Fourier transformation is $\langle {\bf J}\bar{\bf J}\rangle_{\omega} \propto \int \mathrm{d}t\exp(-t/\tau_r+it\omega)$,
and one sees that at $\omega\to 0$ the relaxation time can be estimated in momentum representation as $\tau_r\propto \langle {\bf J}\bar{\bf J}\rangle_{{\bf p}\to 0,\,\omega\to 0}$.
Therefore, the basic contribution to the integral comes from the long wave spectrum part, $p^2\ll |M^2|$, and the relaxation time is
\begin{gather*}
\displaystyle \tau_r\propto\displaystyle \langle {\bf J}\bar{\bf J}\rangle_{{\bf p}\to 0,\,\omega\to 0}\propto \exp\left[ -\dfr{\beta J^2}{ M^2} \right]=\exp\left[ \dfr{bJ^2\beta}{\epsilon g^2\alpha}\dfr{T}{T-T_g} \right].
\end{gather*}
We remind that the long-time correlation of vortices does not mean any long-range ordering of spins themselves, i.e. $\langle \Psi\Psi\rangle|_{r\to\infty} = 0$. This establishes that in frustrated 3D system there is a topological phase transition which is nothing more than glass transition.

\section{Mode-coupling theory equation in the Zwanziger--Mori representation}

The equilibrium value of $|\Psi|^2$ is zero at $T>T_c$ and $|\Psi|^2=\psi^2=\alpha (T_c-T)/Tb$ at $T<T_c$.
We will consider our model in the low temperature interval: $T_g<T<T_c$,  taking into account that the mean magnetization of the system is zero, since the local magnetization deviates from a general direction, $\Psi =\psi +i\delta \Psi$, because of the vortices presence:
\begin{gather*}
\mathcal{H}\approx \dfr{\kappa}{2}{\bf A}\nabla^2{\bf A}-\dfr{M^2}{2}{\bf A}^2+\dfr{\epsilon}2|\nabla \delta\Psi|^2-\dfr{\epsilon}2{g}^2{\bf A}^2|\delta\Psi|^2-\dfr{m^2}2|\delta\Psi|^2+\dfr{b}4|\delta\Psi|^4.
\end{gather*}

In order to describe the non-equilibrium dynamics of the system we will use the Keldysh technics~\cite{Kamenev}.
In terms of this approach the partition function of our system is written as:
\begin{multline*}
\displaystyle \mathcal{Z}=\int D\vec{\bf A}\exp\left[ \beta a^{-d}\tau^{-1}\int\mathrm{d}t\mathrm{d}{\bf r}\left(\mathcal{S}_A+\mathcal{S}_{\Psi}+\epsilon{g}^2\bar{\bf A}{\bf A}|\delta\Psi|^2+\right.\right.\\\left.\left.
\epsilon{g}^2{\bf A}^2\bar{\delta\Psi}\delta\Psi-b\bar{\delta\Psi}{\delta\Psi}^3\right)\right],
\end{multline*}
where $\mathcal{S}_A$ is determined in (\ref{ActionA}), and
\begin{gather*}
\displaystyle \mathcal{S}_{\Psi}={\delta\Psi}\left(\epsilon\nabla^2+\Gamma_{\psi} \tau\partial_t+m^2\right)\bar{\delta\Psi}+\bar{\delta\Psi}\left(\epsilon\nabla^2-
\Gamma_{\psi}\tau\partial_t+m^2\right){\delta\Psi}-2\beta^{-1}{\bar{\delta\Psi}}^2,
\end{gather*}
$\vec {\bf A}=\left\{ \bar {\bf A},\,{\bf A}\right\}$ and $\vec{\delta\Psi}=\left\{ \bar {\delta\Psi},\,{\delta\Psi}\right\}$ are vectors, the components of which are named ``quantum'' and ``classical'' respectively, $\Gamma_A\tau $ and $\Gamma_{\psi}\tau$ are corresponding kinetic coefficients, which correspond to the free gauge field and Higgs field respectively.
The Green functions of the massive vector field components in impulse form present in (\ref{GFA}), and the Green functions of the Higgs field components are:
\begin{multline*}
    G^R_0({\bf p},\,\omega )=\langle \bar{\delta\Psi}{\delta\Psi}\rangle_{{\bf p},\,\omega }=\dfr{\beta^{-1}}{\epsilon{\bf p}^2-m^2- i\Gamma_{\psi}\tau\omega },\\
    G^A_0({\bf p},\,\omega )=\langle{\delta\Psi}\bar{\delta\Psi}\rangle_{{\bf p},\,\omega }=\dfr{\beta^{-1}}{\epsilon{\bf p}^2-m^2+ i\Gamma_{\psi}\tau\omega },\\
    G^K_0({\bf p},\,\omega )=\langle {\delta\Psi\delta\Psi}\rangle_{{\bf p},\,\omega }=\dfr{2\beta^{-2}}{(\epsilon{\bf p}^2-m^2)^2+\Gamma_{\psi}^2\tau
    ^2\omega^2 }.
\end{multline*}

In the framework of the considered functional theory one can easily derive the equation of the mode-coupling theory in the Zwanziger--Mori representation \cite{Vasin,Reichman, Kob} for the dynamic structure factor $S(t)\equiv G_K(t)$. For this, we can use the Dyson equation and the fluctuation--dissipation theorem. We write the Dyson equations for correlation functions $G^A$ and $G^R$:
\begin{gather}\label{GF}
G^{A/R}=G^{A/R}_0+G^{A/R}_0\otimes D^{A/R}\otimes G^{A/R},
\end{gather}
where the symbol $\otimes $ denotes the operation of contraction with respect to $t$ and $\beta^{-1}D^{A(R)}(t)$ is the self-energy part with the physical sense of the memory function.
Let us act on these equations by the operator $\hat G^{-1}_0=\Gamma_{\psi}\tau\partial_t-\epsilon\nabla^2+m^2$ from the left, and then subtract the first expression from the second one:
\begin{gather*}
G^{-1}_0(G^R-G^A)=D^R\otimes G^R-D^A\otimes G^A =
\tau^{-1}\int\limits_0^tD^R(|t'|)G^R(|t-t'|)\mathrm{d}t'.
\end{gather*}
The obtained expression is the equation of the mode-coupling theory in the Zwanziger--Mori representation. To obtain the standard form of this equation, we can include
the time-independent static contribution to the Green function and to the memory function $D$, $D(t)\to D(t)+\tau^{-2}\Omega^2$, where $\Omega $
is the microscopic frequency\cite{Kob}.
Using the fluctuation--dissipation theorem, which implies that $G^R-G^A=\beta\Gamma_{\psi}\tau\partial_t G^K$, we now obtain
\begin{multline*}
\Gamma_{\phi}\tau^2\partial_t^2 G^K(t)+\tau(\epsilon k^2+m^2)\partial_tG^K(t)
+\\ \beta^{-1}\tau^{-2}\Omega^{2}G^K(t)+\beta^{-1}\int\limits_0^tD^R(|t'|)\partial_tG^K(t-t')\mathrm{d}t'=0,
\end{multline*}
where the expression for the memory function is written in the following form (see Appendix I):
\begin{gather}
D^{A/R}(t)\approx \theta(\pm t)\dfr{4a^6(\epsilon g^2)^2\tau}{\pi^2\beta^3\kappa^2\Gamma_A^2\Gamma_{\psi} \sqrt{2\kappa\Gamma_A^{-1}(\kappa\Gamma_A^{-1}+\epsilon\Gamma_{\psi}^{-1})}}
\dfr{e^{-m^2\Gamma_{\psi }^{-1}|t|/\tau}}{|t|}.
\label{DF}
\end{gather}

From (\ref{GF}) and (\ref{DF}) one can estimate the susceptibility time dependence, $\chi(t)$, of the system in the non-ergodic glass state. In this case the correlation function has the following form:
\begin{multline*}
G^{A/R}(t)\approx \theta(\pm t)e^{-m^2\Gamma_{\psi }^{-1}|t|/\tau}\left[1+
\dfr{4a^6(\epsilon g^2)^2}{\pi^2\kappa^2 \sqrt{2\kappa\Gamma_A^{-1}(\kappa\Gamma_A^{-1}+\epsilon\Gamma_{\psi}^{-1})}}
\times\right.\\\left.\dfr{\tau}{\beta^3\Gamma_A^2\Gamma_{\psi}}\int\limits_{0}^{t}\dfr{1}{|t'|}\mathrm{d}t'\right].
\end{multline*}
If the system is in the glass state then $t\ll m^{-2}\Gamma_{\psi }\tau$. Therefore  $\chi(t)=G^A(t)\propto  \ln (t)$. This logarithmic time dependence of susceptibility is the characteristic property of spin glasses.That also confirms us that at $T_g$ the system undergoes the glass transition.

\section{Conclusions}
\label{S3}

We constructed a gauge field theory of the glass transition of the XY-model taking into account formation of vortex topological excitations.
We shown that in case of 3D $\mathrm{XY}$ model the glass transition is the topological phase transition by nature like BKT transition in two-dimensional systems. However, the additional necessary condition for it is the structure frustration, which plays central role at the considered transition. It supplements the thermal fluctuations (which are weak in 3D case) and does not allow system to reach ordered ground state.
Weak frustration shifts a bit the phase transition temperature into the low temperature region.  However, in the case of more hight degree of frustration the system remains in intermediate state (supercooling state) relatively broad temperature interval. In this state the system is locally ordered. It presents a liquid of the vortices. In 3D case the vortices are linear vortices which can move are screening  each other. When the temperature reaches the $T_g$ value the system falls into the disordered confined phase, where the movement of vortices is limited by the frustration and by other vortices (e.g. vortices get entangled).  The effective strength of thermal fluctuations not be able to ``push apart'' the entangled vortices and drive the frustrated system into the ordered state. As a result the system passes into the structurally disordered state of confined vortex lines. It is a disordered state in which there is not a long-range spin correlation. However, the relaxation time diverges in $T=T_g$.  We shown that transition to this non-ordered state is characterised by the Vogel--Fulcher--Tammann law for relaxation time, and has characteristic attributes of glass transition in spin systems. In addition the derived mode-coupling theory equation in the Zwanziger--Mori representation describes the logarithmic time dependence of susceptibility at relaxation that characteristic for spin glasses. As a result one can assert that this is glass state.

\section*{Appendix I}
The memory function has the following form:
\begin{multline*}
\displaystyle D^{A/R}({\bf p},\,t)\approx \dfr{2a^6(\beta\epsilon g^2)^2}{(2\pi)^6}\int \Delta^K_{0}({\bf k+k}_1,\,t)\Delta^K_{0}({\bf k}_1,\,t)G^{A(R)}_0({\bf k+p},\,t)\mathrm{d}^3{\bf k}_1\mathrm{d}^3{\bf k}=\\[12pt]
\displaystyle \theta(\pm t)\dfr{2a^6(\epsilon g^2)^2}{\beta^3(2\pi)^6\Gamma_A^2\Gamma_{\psi}}\iint \dfr{\exp \left[-\left((\kappa k_1^2+M^2)+(\kappa({\bf k}_1+{\bf k})^2+M^2)\right)|t|/{\Gamma_A\tau}\right]}{(\kappa k_1^2+M^2)(\kappa({\bf k}_1+{\bf k})^2+M^2)} \mathrm{d}^3{\bf k}_1\times
\\[12pt]
\displaystyle\exp\left[(\epsilon({\bf k+p})^2-m^2)|t|/{\Gamma_{\psi}\tau} \right]\mathrm{d}^3{\bf k}\approx  \dfr{a^6\theta(\pm t)4\pi (\epsilon g^2)^2e^{(\epsilon p^2-m^2)\Gamma_{\psi }^{-1}|t|/\tau}}{2\pi^4\beta^3\Gamma_A^2\Gamma_{\psi}}\times\\
\int e^{- k^2(\kappa\Gamma_A^{-1}+\epsilon\Gamma_{\psi}^{-1})|t|/\tau}\left[\int\dfr{\exp \left[-2(\kappa k_1^2+M^2)|t|/\Gamma_{A}\tau\right]}
{(\kappa k_1^2+M^2)(\kappa k^2+M^2)} |k_1|^2\mathrm{d}k_1\right]|k|^2\mathrm{d}k.
\end{multline*}
If we consider $M\to 0$, then
\begin{multline*}
\displaystyle D^{A/R}({\bf p},\,t)\approx\dfr{a^6\theta(\pm t)8(\epsilon g^2)^2e^{(\epsilon p^2-m^2)\Gamma_{\psi }^{-1}|t|/\tau}}{\pi^2\beta^3\kappa^2\Gamma_A^2\Gamma_{\psi}}\times\\
\int e^{-k^2(\kappa\Gamma_A^{-1}+\epsilon\Gamma_{\psi}^{-1})|t|/\tau}\mathrm{d}k\int e^{-2\kappa k_1^2\Gamma_{A}^{-1}|t|/\tau}\mathrm{d}k_1.
\end{multline*}
Therefore, at $p\to 0$ we get:
\begin{gather*}
D^{A/R}(t)\approx \theta(\pm t)\dfr{4a^6(\epsilon g^2)^2\tau}{\pi^2\beta^3\kappa^2\Gamma_A^2\Gamma_{\psi} \sqrt{2\kappa\Gamma_A^{-1}(\kappa\Gamma_A^{-1}+\epsilon\Gamma_{\psi}^{-1})}}\dfr{e^{-m^2\Gamma_{\psi }^{-1}|t|/\tau}}{|t|}.
\end{gather*}



\begin{thebibliography}{00}


\bibitem{N1} S.A. Kivelson and G.Tarjus,
Nature Materials 7 (2008) 831.

\bibitem{N2} G.B. McKenna,
Nature Physics 4 (2008) 673.

\bibitem{Tanaka} H. Tanaka, T. Kawasaki, H. Shintani and K. Watanabe,
Nature Materials, Advance Online Publication (2010) 1.

\bibitem{Kibble} T.W.B. Kibble,
J.\,Phys.\,A:\,Math.\,Gen. 9 (1976) 1387.

\bibitem{Zurek} W.H. Z\"{u}rek,
Nature 317 (1985) 505.

\bibitem{Brazhkin1} V.V. Brazhkin, Yu.D. Fomin, A.G. Lyapin, V.N. Ryzhov and K. Trachenko,
Phys. Rev. E 85 (2012) 031203.

\bibitem{Brazhkin2} Yu.D. Fomin, V.N. Ryzhov, E.N. Tsiok, and V.V. Brazhkin,
Scientific Reports  5 (2015) 14234.



\bibitem{Ber} V.L. Berezinskii,
Sov. Phys.--JETP 32 (1970) 493.

\bibitem{KT1972} J.M. Kosterlitz, \& D.J. Thouless
Journal of Physics C: Solid State Phys. 5 (1972) L124.

\bibitem{KostThoul} J.M. Kosterlitz, \& D.J. Thouless,
J.\,Phys. C: Solid State Phys.  6 (1973) 1181.

\bibitem{Kosterlitz1977}
J.M. Kosterlitz,
Journal of Physics C: Solid State Phys. 7, (1974) 1046.

\bibitem{Jose1977}
J. Jos\'{e}, L.P. Kadanoff, S. Kirkpatric \& D.R. Nelson,
Phys. Rev. B 16 (1977); Errata in Phys. Rev. B 17 (1978) 1477.

\bibitem{Anderson1978}
P.W. Anderson,
\textit{Lectures on amorphous systems}, in Les Houches, Session XXXI, 1978 - La mati\`{e}re mal condens\'{e}e/III-condensed matter, 1978, edited by R. Balian et al. (North-Holland, Amsterdam, 1978).

\bibitem{VV} M.G. Vasin, and V.M. Vinokur,
Physics of the solid state 60, No. 6 (2018) 1195.

\bibitem{F1} E. Fradkin, L. Susskind,
Phys. Rev. D 17 (1978) 2637.
\bibitem{F2} E. Fradkin, B.A. Huberman, S.H. Shenker 
Phys. Rev. B 18 (1978) 4789.

\bibitem{R37} I.E. Dzyaloshinskii, and G.E. Volovik, Ann. Phys. (1980).
\bibitem{R38} I.E. Dzyaloshinskii, and S.P. Obukhov,
Sov. Phys. JETP 56 (1982).
\bibitem{R22} G. Toulouse,
Comm. Physics 2 (1977) 115.
\bibitem{R39} J. Villain,
J. Phys. C 11 (1978) 745.
\bibitem{R23} N. Rivier, D.M. Duffy,
J. Physique 43 (1982) 293.
\bibitem{R34} D. Kivelson, G. Tarjus,
Phyl. Mag. B 77 (1998) 245.
\bibitem{R40} D.R. Nelson,
Phys. Rev. B 28 (1983) 5515.
\bibitem{R42} B.I. Halperin, Statistical Mechanics of Topological Defects, in: Les Houches. Session XXXV. Physics of Defects. Ed. R. Balian, M. Kleman, and J.-P. Poirier. North-Holland Publishing Company (1981).
\bibitem{Obukhov} S.P. Obukhov,
Zh. Eksp. Teor. Fiz. 83 (1982) 1978.





\bibitem{DV5} I.E. Dzyaloshinskii, \& G.E. Volovik,
J. de Physique  39 (1978) 693.




\bibitem{Hertz} J.A. Hertz,
Phys. Rev. B 18 (1978) 4875.

\bibitem{R44} G. Kohring, R.E. Shrock, \& P. Wills,
 Phys. Rev. Lett. 57 (1986)  1358.

\bibitem{Olsson1} P. Olsson,
Phys. Rev. B 52 (1995)  4526.

\bibitem{Olsson2} P. Olsson,
Phys. Rev. Lett. 91 (2003) 077002.

\bibitem{R45} M. Camarda, F. Siringo, R. Pucci, A. Sudbø, \& J. Hove,
Phys. Rev. B 74 (2006)  104507.

\bibitem{R46} D.A. Garanin, E.M. Chudnovsky, \& T. Proctor,
Phys. Rev. B 88 (2013)  224418.

\bibitem{Polyakov}
A.M. Polyakov, \textit{Gauge Fields and Strings} (Harwood Academic Publishers, Chur, Switzerland, 1987).

\bibitem{R2} N. Rivier,
Revista Brasileira de Flsica 15,  4 (1985) 311.




\bibitem{Min} P. Minnhagen,
Reviews of Modern Physics 59 (1987) 1001.

\bibitem{Zee}
 A. Zee, \textit{Quantum Field Theory in a Nutshell} (Princeton University Press, Princeton, 2010) ISBN\,9780691140346.

\bibitem{Cvel}
A.M. Tsvelik, \textit{Quantum Field Theory in Condensed Matter Physics} (Cambridge University Press, Cambridge, 1998) ISBN\,0521589894.

\bibitem{binder} K. Binder, \& A.P. Young,
Reviews of Modern Physics 58 (1986) 801.



\bibitem{K2} G. Tarjus, S.A. Kivelson, Z. Nussinov, \& P. Viot,
J. Phys: Cond. Matter  17 (2005)  R1143.


\bibitem{Nussinov} Z. Nussinov,
Phys. Rev. B 69 (2004)   014208.

\bibitem{Vasin}
M.G. Vasin,
Journal of Statistical Mechanics: Theory and Experiment (2011)  P05009.

\bibitem{SupVasin}
M.G. Vasin,
Physica A 415 (2014) 533.

\bibitem{Kamenev} A. Kamenev,
Field theory of non-equilibrium systems. \textit{Cambridge University Press, New York} ISBN\,9780521760829 (2011).

\bibitem{HH} P.C. Hohenberg, \& B.I. Halperin,
Reviews of Modern Physics  49, 3 (1977) 435.

\bibitem{Reichman} D.R. Reichman and P. Charbonneau, J.\,Stat.\,Mech. (2005) P05013.

\bibitem{Kob} W. Kob, The Mode-Coupling Theory of the Glass Transition // Experimental and Theoretical Approaches to Supercooled Liquids: Advances and Novel Applications Eds.: J. Fourkas, D. Kivelson, U. Mohanty, and K. Nelson (ACS Books, Washington. -- 1997.

\bibitem{PP} A.Z. Patashinskii, \& V.L. Pokrovskii, \textit{Fluctuation Theory of Phase Transitions} (Pergamon Press, Oxford, 1979) ISBN\,0080216641.









\end{thebibliography}


\section*{Acknowledgements}

The work was supported 
by Russian Foundation for Basic Research, Grants 18-02-00643 (MV).
I am grateful to Valeriy Vinokur for stimulating discussions.

\end{document}